\documentclass[aps,pra]{revtex4}
\usepackage{graphicx}

\begin{document}

\title{Faithful and deterministic teleportation of an arbitrary $N$-qubit state
using and identifying only Bell states
\thanks{E-mail: Zhangzj@wipm.ac.cn} }

\author{Zhan-jun Zhang$^{a,b}$ \\
{\footnotesize  $^a$School of Physics \& Material Science, Anhui
University, Hefei 230039, China} \\
{\footnotesize  $^b$Wuhan Institute of Physics and Mathematics,
The Chinese Academy of Sciences, Wuhan 430071, China} \\
{\footnotesize E-mail: zhangzj@wipm.ac.cn }}

\date{\today}
\maketitle

\begin{minipage}{420pt}
First, I show explicitly a scheme to {\it faithfully} and {\it
deterministically} teleport an arbitrary 2-qubit state from Alice
to Bob. In this scheme two same Bell states are sufficient for
use. Bob can recover the 2-qubit state by performing at most 4
single-qubit unitary operations conditioned on Alice's 4-bit
classical public message corresponding to her two Bell-state
measurement outcomes. Then I generalize the 2-qubit teleportation
scheme to an aribitrary $N(N\ge 3)$-qubit state teleportation case
by using $N$ same Bell states. In the generalized scheme, Alice
only needs to identify $N$ Bell states after quantum swapping and
then publish her measurement outcomes ($2N$-bit classical
message). Conditioned on Alice's $2N$-bit classical message, Bob
only needs to perform at most $2N$ single-qubit unitary operations
to {\it fully} recover the arbitrary state. By comparing with the
newest relevant work [Phys. Rev. A{\bf 71}, 032303(2005)], the
advantages of the present schemes are revealed, respectively.\\

\noindent {\it PACS: 03.67.Hk, 03.65.Ud} \\
\end{minipage}\\

{\bf 1. Introduction}\\

Since no-cloning theorem forbids a copy of an arbitrary unknown
quantum state, how to interchange different resources has ever
been a question in quantum computation and quantum information. In
1993, Bennett {\it et al}[1] first proposed a method teleporting
an arbitrary unknown quantum state in a qubit to a distant qubit
with the aid of Einstein-Podlsky-Rosen (EPR) pair. Their work
showed in essence the interchangeability of different resources in
quantum mechanics. Hence, after Bennett {\it et al}'s poineering
work[1], quantum teleportation has attracted many attentions in
both theoretical and experimental aspects[2-18].

In theoretical aspect, one important question was whether it was
possible to teleport not just a single qubit, but rather
$N(N\ge2)$ qubits. To our knowledge, so far there have been
several explorations. For examples, Gao {\it et al} used a kind of
generalized Bell states to realize a {\it probabilistic} two-qubit
teleportation [15]; Fang {\it et al} presented a scheme which can
{\it probabilistically} teleport a three-particle state via three
pairs of entangled particles [12]; Yang and Guo proposed a quantum
teleportation scheme of a {\it special} Greenberger-Horne-Rosen
(GHZ) state [6]; In [8] Lee {\it et al} presented a scheme to
teleport a {\it special} classes of two-qubit states; and so on.
Later, Lee {\it et al} in a very interesting work, showed that it
was possible to teleport an {\it arbitrary} two-qubit state from
Alice to Bob by {\it using a four-entangled state} and sending to
him four bits of classical information [9]. Due to Lee {\it et
al}'s inexplicit expression of their latter scheme, very recently
Rigolin has revisited the scheme and generalized it to a $N(N\ge
3)$-qubit case[16]. In his work, Rigolin has explicitly presented
the {\it faithful and deterministic} quantum teleportation scheme
of an {\it arbitrary} two-qubit state and established its relation
to multipartite entanglement. This is an important and interesting
progress.

However, I  think, there lie two disadvantages in Rigolin's this
work[16]: (1) A complicated entangled state (e.g., one of the
so-called generalized Bell states in his equations 4-19) is
necessary for use. Although according to the present-day
technologies it is possible to synthesize a generalized Bell state
[16] or a multi-photon GHZ state from Bell states [19], the total
production efficiency is lower for the produce is
nondeterministic. Moreover, in fact an experimental synthesization
is very difficult, especially when the photon amount in a
generalized Bell state or a GHZ state is large. Incidentally, to
my knowledge, so far only up to five-photon GHZ state has been
synthesized[19]. Hence, it will be very nice if one can eliminate
the synthesization process, i.e., directly using the same amount
of Bell state instead of a generalized Bell states synthesized
from these Bell states, for in this case the use of the
entanglement resource is more economical and corresponding
experimental prepare difficulties can be greatly reduced. (2) All
the generalized Bell states must be successfully discriminated. It
is generally admitted that an identification of a Bell state
should be much easier than an identification of a generalized Bell
state or a GHZ state, though a complete recognition of Bell states
is far to come. Hence, it will also be very nice if one only needs
to discriminate Bell states instead of generalized Bell states or
GHZ states. In this case, obviously, the experimental
discrimination difficulty is reduced. To eliminate the two
disadvantages, in this paper I will present an arbitrary $N(N\ge
2)$-qubit state quantum teleportation scheme by using and
identifying only Bell states. Within my scheme an {\it arbitrary}
$N(N\ge 2)$-qubit state can also be {\it faithfully} and {\it
deterministically} teleported from Alice to Bob.

The paper is organized as follows. In section 2, I will present a
specific quantum teleportation scheme of an arbitrary 2-qubit
state, and a comparison between this scheme and the counterpart
scheme in [16] is made. Then in section 3, I will generalize the
2-qubit teleportation scheme to a $N(N\ge 3)$-qubit teleportation
case also by using and identifying only Bell states.
A concluding summary is given in section 4.\\

{\bf 2. Quantum teleportation scheme of an arbitrary 2-qubit
state}\\

Before giving my specific quantum teleportation scheme of an
arbitrary 2-qubit state, let us briefly review the quantum
teleportation scheme of an arbitrary single-qubit state, which is
originally proposed by Bennett et al in 1993[1]. The scheme is
executed as follows. Alice and Bob initially share a Bell state
(i.e., a maximally entangled two-qubit state), say,
$|\Psi^-\rangle$. By the way, the four Bell states are defined as
follows
\begin{eqnarray}
|\Phi^{\pm}\rangle=(|00\rangle \pm |11\rangle)/\sqrt{2},\\
|\Psi^{\pm}\rangle=(|01\rangle \pm |10\rangle)/\sqrt{2}.
\end{eqnarray}
Then the state of the joint system, i.e., the entity of the
arbitrary state $|u\rangle=\alpha|0\rangle+\beta|1\rangle$ to be
teleported and the shared Bell state, can be written as
\begin{eqnarray}
|\varphi\rangle= |u\rangle_x \otimes |\Psi^-\rangle_{ab},
\end{eqnarray}
where $x$ stands for the arbitrary qubit and the subscripts $a$
and $b$ in the Bell state $|\Psi^-\rangle$ label the qubits
belonging to Alice and Bob, respectively. The equation 3 can be
rewritten as follow,
\begin{eqnarray}
|\varphi\rangle=\frac{1}{2}\{ |\Psi^-\rangle_{xa}
(-\alpha|0\rangle_b-\beta|1\rangle_b)+|\Psi^+\rangle_{xa}
(-\alpha|0\rangle_b+\beta|1\rangle_b) \nonumber \\
+|\Phi^-\rangle_{xa}(\alpha|1\rangle_b+\beta|0\rangle_b)+|\Phi^+\rangle_{xa}
(\alpha|1\rangle_b-\beta|0\rangle_b)\}.
\end{eqnarray}
Alice now performs a Bell state measurement on her two qubits and
classically communicates the measurement outcome to Bob.
Conditioned on Alice's two-bit information, Bob can carry out the
appropriate unitary operation to reconstruct the unknown arbitrary
state in his qubit $b$. See Table 1. \\

Table 1 Alice's Bell-state measurement outcome, Bob's
corresponding unitary operation and the final state in his qubit.
$I$ is an identity operator and $\sigma$ are the usual Pauli
matrices. $\sigma^z|1\rangle=|1\rangle$,
$\sigma^z|0\rangle=-|0\rangle$; $\sigma^x|1\rangle=|0\rangle$,
$\sigma^x|0\rangle=|1\rangle$.
\begin{center}
\begin{tabular}{ccccc} \hline
Alice's Bell-state &&  Bob's corresponding && The final state in \\
measurement outcome &&  unitary operation && Bob's qubit \\ \hline
$|\Psi^-\rangle_{xa}$ && $I_b$&&
$-\alpha|0\rangle_b-\beta|1\rangle_b=-|u\rangle_b$ \\
$|\Psi^+\rangle_{xa}$ && $\sigma_b^z$&& $\alpha|0\rangle_b+\beta|1\rangle_b=|u\rangle_b$\\
$|\Phi^-\rangle_{xa}$ && $\sigma_b^x$ &&$\alpha|0\rangle_b+\beta|1\rangle_b=|u\rangle_b$\\
$|\Phi^+\rangle_{xa}$&& $i\sigma^y_b=\sigma^z_b\sigma^x_b$ &&
$-\alpha|0\rangle_b-\beta|1\rangle_b=-|u\rangle_b$\\ \hline
\end{tabular} \\
\end{center}

\vskip 1cm Now let us first present our specific scheme of
teleporting two arbitrary qubits by using and identifying only
Bell states. Suppose that the arbitrary 2-qubit state Alice wants
to teleport to Bob is written as
\begin{eqnarray}
|\phi\rangle_{x_1x_2}=\alpha|0\rangle_{x_1}|0\rangle_{x_2}+\beta|0\rangle_{x_1}|1\rangle_{x_2}
+\gamma|1\rangle_{x_1}|0\rangle_{x_2}+\delta|1\rangle_{x_1}|1\rangle_{x_2},
\end{eqnarray}
where $x_1$ and $x_2$ label the two arbitrary qubits in Alice's
site, $\alpha$,$\beta$,$\gamma$ and $\delta$ are {\it unknown}
complex coefficients and $|\phi\rangle_{x_1x_2}$ is assumed to be
normalized. Alice and Bob must share in advance a pair of same
Bell states, say, $\Psi^-_{a_2b_2} \otimes \Psi^-_{a_1b_1}$.
Incidentally, the two qubits $b_1$ and $b_2$ in Bob's site are
used to "receive" the teleported state from Alice. Hence, the
initial joint state is
\begin{eqnarray}
|\psi\rangle_{x_1x_2a_2b_2a_1b_1}
=|\phi\rangle_{x_1x_2}\otimes\Psi^-_{a_2b_2}\otimes
\Psi^-_{a_1b_1}.
\end{eqnarray}
It can be rewritten as
\begin{eqnarray}
&& |\psi\rangle_{x_1x_2a_2b_2a_1b_1} \nonumber \\ &=&
[|0\rangle_{x_1}(\alpha|0\rangle_{x_2}+\beta|1\rangle_{x_2})+
|1\rangle_{x_1}(\gamma|0\rangle_{x_2}+\delta|1\rangle_{x_2})]\otimes\Psi^-_{a_2b_2}\otimes
\Psi^-_{a_1b_1} \nonumber \\
&=& \frac{1}{2}\{|0\rangle_{x_1}[- |\Psi^-\rangle_{x_2a_2}
  I(\alpha|0\rangle_{b_2}+\beta|1\rangle_{b_2})+
|\Psi^+\rangle_{x_2a_2}\sigma_{b_2}^z(\alpha|0\rangle_{b_2}+\beta|1\rangle_{b_2})\nonumber
\\
&&+|\Phi^-\rangle_{x_2a_2}
I(\alpha|1\rangle_{b_2}+\beta|0\rangle_{b_2})+|\Phi^+\rangle_{x_2a_2}\sigma_{b_2}^x\sigma_{b_2}^z
(\alpha|0\rangle_{b_2}+\beta|0\rangle_{b_2})]\nonumber
\\ &+& |1\rangle_{x_1}[-|\Psi^-\rangle_{x_2a_2}
  (\gamma|0\rangle_{b_2}+\delta|1\rangle_{b_2})+|\Psi^+\rangle_{x_2a_2}\sigma_{b_2}^z
(\gamma|0\rangle_{b_2}+\delta|1\rangle_{b_2})]\nonumber\\
&&+|\Phi^-\rangle_{x_2a_2}
I(\gamma|1\rangle_{b_2}+\delta|0\rangle_{b_2})+|\Phi^+\rangle_{x_2a_2}
\sigma_{b_2}^x\sigma_{b_2}^z(\gamma|0\rangle_{b_2}+\delta|1\rangle_{b_2})]
\}\otimes\Psi^-_{a_1b_1} \nonumber \\
&=& \frac{1}{2}\{|0\rangle_{x_1}[(-|\Psi^-\rangle_{x_2a_2}
  I+|\Psi^+\rangle_{x_2a_2}\sigma_{b_2}^z+|\Phi^-\rangle_{x_2a_2}
I+|\Phi^+\rangle_{x_2a_2}\sigma_{b_2}^x\sigma_{b_2}^z)
(\alpha|0\rangle_{b_2}+\beta|0\rangle_{b_2})]\nonumber
\\ &+& |1\rangle_{x_1}[(-|\Psi^-\rangle_{x_2a_2}
+|\Psi^+\rangle_{x_2a_2}\sigma_{b_2}^z +|\Phi^-\rangle_{x_2a_2}
I+|\Phi^+\rangle_{x_2a_2}
\sigma_{b_2}^x\sigma_{b_2}^z)(\gamma|0\rangle_{b_2}+\delta|1\rangle_{b_2})]
\}\otimes\Psi^-_{a_1b_1}.
\end{eqnarray}
Define that
\begin{eqnarray}
O_{x_2a_2;b_2}\equiv|\Psi^-\rangle_{x_2a_2}
  I+|\Psi^+\rangle_{x_2a_2}\sigma_{b_2}^z+|\Phi^-\rangle_{x_2a_2}
\sigma_{b_2}^x+|\Phi^+\rangle_{x_2a_2}\sigma_{b_2}^x\sigma_{b_2}^z,
\end{eqnarray}
then the equation 7 can be written as
\begin{eqnarray}
&& |\psi\rangle_{x_1x_2a_2b_2a_1b_1}\nonumber \\&=&
\frac{1}{2}[|0\rangle_{x_1}O_{x_2a_2;b_2}
(\alpha|0\rangle_{b_2}+\beta|0\rangle_{b_2}) +
|1\rangle_{x_1}O_{x_2a_2;b_2}(\gamma|0\rangle_{b_2}+\delta|1\rangle_{b_2})
]\otimes\Psi^-_{a_1b_1}\nonumber \\ &=& \frac{1}{4}\{
|\Psi^-\rangle_{x_1a_1}
  [-|0\rangle_{b_1}O_{x_2a_2;b_2}(\alpha|0\rangle_{b_2}+\beta|1\rangle_{b_2})-
  |1\rangle_{b_1}O_{x_2a_2;b_2}(\gamma|0\rangle_{b_2}+\delta|1\rangle_{b_2})] \nonumber\\
&&+ |\Psi^+\rangle_{x_1a_1}
  [-|0\rangle_{b_1}O_{x_2a_2;b_2}(\alpha|0\rangle_{b_2}+\beta|1\rangle_{b_2})+
  |1\rangle_{b_1}O_{x_2a_2;b_2}(\gamma|0\rangle_{b_2}+\delta|1\rangle_{b_2})] \nonumber\\
&&+|\Phi^-\rangle_{x_1a_1}
  [|1\rangle_{b_1}O_{x_2a_2;b_2}(\alpha|0\rangle_{b_2}+\beta|1\rangle_{b_2})+
  |0 \rangle_{b_1}O_{x_2a_2;b_2}(\gamma|0\rangle_{b_2}+\delta|1\rangle_{b_2})] \nonumber\\
&&+|\Phi^-\rangle_{x_1a_1}
  [|1\rangle_{b_1}O_{x_2a_2;b_2}(\alpha|0\rangle_{b_2}+\beta|1\rangle_{b_2})-
  |0 \rangle_{b_1}O_{x_2a_2;b_2}(\gamma|0\rangle_{b_2}+\delta|1\rangle_{b_2})]\}\nonumber \\
&=& \frac{1}{4}\{ -|\Psi^-\rangle_{x_1a_1}O_{x_2a_2;b_2}
  [|0\rangle_{b_1}(\alpha|0\rangle_{b_2}+\beta|1\rangle_{b_2})+
  |1\rangle_{b_1}(\gamma|0\rangle_{b_2}+\delta|1\rangle_{b_2})] \nonumber\\
&&+ |\Psi^+\rangle_{x_1a_1}O_{x_2a_2;b_2}\sigma_{b_1}^z
  [|0\rangle_{b_1}(\alpha|0\rangle_{b_2}+\beta|1\rangle_{b_2})+
  |1\rangle_{b_1}(\gamma|0\rangle_{b_2}+\delta|1\rangle_{b_2})] \nonumber\\
&&+|\Phi^-\rangle_{x_1a_1}O_{x_2a_2;b_2}\sigma_{b_1}^x
  [|0\rangle_{b_1}(\alpha|0\rangle_{b_2}+\beta|1\rangle_{b_2})+
  |1\rangle_{b_1}(\gamma|0\rangle_{b_2}+\delta|1\rangle_{b_2})] \nonumber\\
&&+|\Phi^-\rangle_{x_1a_1}O_{x_2a_2;b_2}\sigma_{b_1}^x\sigma_{b_1}^z
  [|0\rangle_{b_1}(\alpha|0\rangle_{b_2}+\beta|1\rangle_{b_2})-
  |1\rangle_{b_1}(\gamma|0\rangle_{b_2}+\delta|1\rangle_{b_2})]
\}\nonumber \\
&=& \frac{1}{4}( -|\Psi^-\rangle_{x_1a_1}O_{x_2a_2;b_2}
  + |\Psi^+\rangle_{x_1a_1}O_{x_2a_2;b_2}\sigma_{b_1}^z +|\Phi^-\rangle_{x_1a_1}O_{x_2a_2;b_2}\sigma_{b_1}^x
\nonumber \\ &&
+|\Phi^-\rangle_{x_1a_1}O_{x_2a_2;b_2}\sigma_{b_1}^x\sigma_{b_1}^z)
|\phi\rangle_{b_1b_2} .
\end{eqnarray}

Alice performs Bell-state measurements on the qubit pairs
$(x_1,a_1)$ and $(x_2,a_2)$ in her site. With equal probabilities
(1/16) she obtains a Bell-state pair (see the equation 9 and Table
2). Then she sends to Bob a classical message of four bits (each
Bell state corresponds to a classical message of two bits) to
inform him her measurement outcomes. With this information Bob
knows what single-qubit unitary operations (see Table 2) he must
apply on his two qubits $b_1$ and $b_2$ to recover correctly the
teleported state $|\phi\rangle$. After Bob's unitary operations
the scheme is completed and Alice's arbitrary two-qubit state has
been successfully teleported to Bob.

Let us simply compare the present scheme with the counterpart
scheme in Ref.[16] as follows. (1) In the present 2-qubit quantum
teleportation scheme only two same Bell states (e.g., $\Psi^-
\otimes \Psi^-$) are sufficient for use, while in the 2-qubit
quantum teleportation scheme in Ref.[16] a generalized Bell state
(e.g.,
$|g_1\rangle=\frac{1}{2}(|0000\rangle+|0101\rangle+|1010\rangle+|1111\rangle)$)
should be prepared for use. Two Bell states can not be
deterministically synthesized to a generalized Bell state, hence a
synthesization of a generalized Bell state should need more than 2
Bell states on average. Moreover, if two Bell states instead of a
generalized Bell state are used, then the experimental
synthesization process is not needed anymore. Obviously, in both
the aspect of the entanglement resource use and the aspect of
experiment, it is economical to use two Bell states directly
instead of a generalized Bell state. (2) In the present 2-qubit
quantum teleportation scheme only the four Bell states defined in
eqs.(1-2) need to be discriminated. In contrast, within the
counterpart scheme in Rigolin's work [16] all 16 generalized Bell
states defined by the eqs.(4-19) in Ref.[16] should be
discriminated. Obviously, the experimental recognition
difficulties in the present scheme are considerably reduced. (3)
In both schemes, Alice sends a 4-bit classical message
corresponding to the measurement outcome(s) to Bob. (4)In both
schemes, conditioned on Alice's message, Bob needs to perform at
most 4 single-qubit unitary operations to fully recover the
arbitrary state. Incidentally the controlled-NOT gate is also not
necessary in this scheme as well as in [16]. Hence, it is obvious
that the present scheme overwhelms the counterpart scheme in [16].\\

Table 2 Alice's Bell-state measurement outcomes and their
probabilities, and Bob's corresponding unitary operations
conditioned on Alice's measurement outcomes.
\begin{center}
\begin{tabular}{ccccc} \hline
Alice's Bell-state && probability &&  Bob's corresponding \\
measurement outcomes && && unitary operations \\ \hline
$|\Psi^-\rangle_{x_2a_2},|\Psi^-\rangle_{x_1a_1}$ && 1/16 &&$I^{-1}=I$\\
$|\Psi^-\rangle_{x_2a_2},|\Psi^+\rangle_{x_1a_1}$ && 1/16 &&$(\sigma^z_{b_1})^{-1}=\sigma^z_{b_1}$\\
$|\Psi^-\rangle_{x_2a_2},|\Phi^-\rangle_{x_1a_1}$ && 1/16 &&$(\sigma^x_{b_1})^{-1}=\sigma^x_{b_1}$\\
$|\Psi^-\rangle_{x_2a_2},|\Phi^+\rangle_{x_1a_1}$ && 1/16 &&$(\sigma^z_{b_1}\sigma^x_{b_1})^{-1}=\sigma^x_{b_1}\sigma^z_{b_1}$\\
$|\Psi^+\rangle_{x_2a_2},|\Psi^-\rangle_{x_1a_1}$ && 1/16 &&$(\sigma^z_{b_2})^{-1}=\sigma^z_{b_2}$\\
$|\Psi^+\rangle_{x_2a_2},|\Psi^+\rangle_{x_1a_1}$ && 1/16 &&$(\sigma^z_{b_2}\sigma^z_{b_1})^{-1}=\sigma^z_{b_1}\sigma^z_{b_2}$\\
$|\Psi^+\rangle_{x_2a_2},|\Phi^-\rangle_{x_1a_1}$ && 1/16 &&$(\sigma^z_{b_2})^{-1}=\sigma^z_{b_2}$\\
$|\Psi^+\rangle_{x_2a_2},|\Phi^+\rangle_{x_1a_1}$ && 1/16 &&
$(\sigma^z_{b_2}\sigma^z_{b_1}\sigma^x_{b_1})^{-1}=\sigma^x_{b_1}\sigma^z_{b_1}\sigma^z_{b_2}$\\
$|\Phi^-\rangle_{x_2a_2},|\Psi^-\rangle_{x_1a_1}$ && 1/16 &&
$(\sigma_{b_2}^x\sigma_{b_1}^x)^{-1}=\sigma_{b_1}^x\sigma_{b_2}^x$\\
$|\Phi^-\rangle_{x_2a_2},|\Psi^+\rangle_{x_1a_1}$ && 1/16 &&
$(\sigma_{b_2}^x\sigma_{b_1}^x\sigma_{b_1}^z)^{-1}=\sigma_{b_1}^z\sigma_{b_1}^x\sigma_{b_2}^x$\\
$|\Phi^-\rangle_{x_2a_2},|\Phi^-\rangle_{x_1a_1}$ && 1/16 &&$(\sigma_{b_2}^x)^{-1}=\sigma_{b_2}^x$\\
$|\Phi^-\rangle_{x_2a_2},|\Phi^+\rangle_{x_1a_1}$ && 1/16 &&
$(\sigma^x_{b_2}\sigma_{b_1}^z)^{-1}=\sigma_{b_1}^z\sigma^x_{b_2}$\\
$|\Phi^+\rangle_{x_2a_2},|\Psi^-\rangle_{x_1a_1}$ && 1/16 &&
$(\sigma_{b_2}^x\sigma_{b_2}^z\sigma^x_{b_1})^{-1}=\sigma_{b_1}^x\sigma_{b_2}^z\sigma^x_{b_2}$\\
$|\Phi^+\rangle_{x_2a_2},|\Psi^+\rangle_{x_1a_1}$ && 1/16 &&
$(\sigma_{b_2}^x\sigma^z_{b_2}\sigma_{b_1}^x\sigma^z_{b_1})^{-1}
=\sigma^z_{b_1}\sigma_{b_1}^x\sigma^z_{b_2}\sigma_{b_2}^x$\\
$|\Phi^+\rangle_{x_2a_2},|\Phi^-\rangle_{x_1a_1}$ && 1/16 &&
 $(\sigma_{b_2}^x\sigma^z_{b_2})^{-1}=\sigma_{b_2}^z\sigma^x_{b_2}$\\
$|\Phi^+\rangle_{x_2a_2},|\Phi^+\rangle_{x_1a_1}$ && 1/16 &&
$(\sigma^z_{b_2}\sigma_{b_2}^x\sigma^z_{b_1})^{-1}
=\sigma^z_{b_1}\sigma_{b_2}^x\sigma^z_{b_2}$\\
\hline
\end{tabular} \\
\end{center}
\vskip 1cm

{\bf 3. Quantum teleportation scheme of an arbitrary
$N(N\ge3)$-qubit state} \\

Now let us generalize the arbitrary 2-qubit state quantum
teleportation scheme to an arbitrary $N(N\ge 3)$-qubit quantum
state teleportation scheme. In the following I will prove that, if
an arbitrary $(N-1)(N\ge 3)$-qubit quantum state can be teleported
successfully between Alice and Bob via sharing $N-1$ same Bell
states, identifying $N-1$ Bell states after quantum swapping,
sending $2(N-1)$ bits information and performing at most $2(N-1)$
single-qubit operations, then an arbitrary $N(N\ge 3)$-qubit
quantum state can be teleported successfully via sharing $N$ same
Bell states, identifying $N$ Bell states after quantum swapping,
sending $2N$ bits information and performing at most $2N$
single-qubit operations.

Suppose that the arbitrary $n(n\ge 3)$-qubit state Alice wants to
teleport to Bob is written as
\begin{eqnarray}
|\xi\rangle_{x_1x_2\dots x_N}= \sum\limits_{m_N=0}^1\dots
\sum\limits_{m_2=0}^1\sum\limits_{m_1=0}^1C_{m_1m_2\dots
m_N}|m_1\rangle_{x_1}|m_2\rangle_{x_2}\dots|m_N\rangle_{x_N},
\end{eqnarray}
where $C$'s are complex coefficients and $|\xi\rangle_{x_1x_2\dots
x_N}$ is assumed to be normalized. It can be decomposed as
\begin{eqnarray}
|\xi\rangle_{x_1x_2\dots x_N}&=&
|0\rangle_{x_N}(\sum\limits_{m_{N-1}=0}^1\dots
\sum\limits_{m_2=0}^1\sum\limits_{m_1=0}^1C_{m_1m_2\dots
m_{N-1}0}|m_1\rangle_{x_1}|m_2\rangle_{x_2}\dots|m_{N-1}\rangle_{x_{N-1}})
\nonumber \\ &+& |1\rangle_{x_N}(\sum\limits_{m_{N-1}=0}^1\dots
\sum\limits_{m_2=0}^1\sum\limits_{m_1=0}^1C_{m_1m_2\dots
m_{N-1}1}|m_1\rangle_{x_1}|m_2\rangle_{x_2}\dots|m_{N-1}\rangle_{x_{N-1}})\nonumber\\
&\equiv& |0\rangle_{x_N}\zeta_{x_1x_2\dots
x_{N-1}}+|1\rangle_{x_N}\zeta'_{x_1x_2\dots x_{N-1}}.
\end{eqnarray}
Here $\zeta_{x_1x_2\dots x_{N-1}}$ and $\zeta'_{x_1x_2\dots
x_{N-1}}$ are in essence arbitrary $(N-1)$-qubit states,
respectively. Note that they have the same form except for their
coefficients. This is important for the later use.

Alice and Bob must share in advance $N$ same Bell states, say,
$\Psi^-_{a_Nb_N} \otimes \dots \otimes \Psi^-_{a_2b_2} \otimes
\Psi^-_{a_1b_1}$. Similarly, as mentioned before,  the $N$ qubits
$b_1$, $b_2$, \dots, $b_{N-1}$and $b_N$ in Bob's site are used to
"receive" the teleported state from Alice. Hence, the initial
joint state is
\begin{eqnarray}
&& |\chi\rangle_{x_1x_2\dots x_Na_Nb_N\dots a_2b_2a_1b_1}
\nonumber \\&=&|\xi\rangle_{x_1x_2\dots
x_N}\otimes\Psi^-_{a_Nb_N}\otimes\dots\otimes\Psi^-_{a_2b_2}\otimes
\Psi^-_{a_1b_1} \nonumber \\ &=&
[|0\rangle_{x_N}(\zeta_{x_1x_2\dots
x_{N-1}}\otimes\Psi^-_{a_{N-1}b_{N-1}}\otimes\dots\otimes\Psi^-_{a_2b_2}\otimes
\Psi^-_{a_1b_1})\nonumber \\
&+& |1\rangle_{x_N}(\zeta'_{x_1x_2\dots x_{N-1}}
\otimes\Psi^-_{a_{N-1}b_{N-1}}\otimes\dots\otimes\Psi^-_{a_2b_2}\otimes
\Psi^-_{a_1b_1}) ]\otimes\Psi^-_{a_Nb_N}.
\end{eqnarray}
Since I have previously supposed that an arbitrary $(N-1)(N\ge
3)$-qubit quantum state can be teleported successfully via using
and identifying only Bell states, then the following equation
holds,
\begin{eqnarray}
& & \zeta_{x_1x_2\dots
x_{N-1}}\otimes\Psi^-_{a_{N-1}b_{N-1}}\otimes\dots\otimes\Psi^-_{a_2b_2}\otimes
\Psi^-_{a_1b_1} \nonumber \\ &=&
\frac{1}{2^{N-1}}\sum\limits_{i=1}^{2^{N-1}}{\cal
H}_{i;x_{N-1}a_{N-1}}\dots {\cal P} _{i;x_2a_2}{\cal Q}_{i;x_1a_1}
U_{i;b_{N-1}\dots b_2b_1}\zeta_{b_1b_2\dots b_{N-1}}\nonumber \\
&\equiv& \frac{1}{2^{N-1}}\sum\limits_{i=1}^{2^{N-1}}
O_{i}\zeta_{b_1b_2\dots b_{N-1}},
\end{eqnarray}
where each calligraphic letter stands for a Bell state and $U$ is
a unitary operation containing at most $2(N-1)$ single-qubit
operations, hence each $O$ contains $N-1$ Bell states and at most
$2(N-1)$ single-qubit operations. By the way, since
$\zeta_{x_1x_2\dots x_{N-1}}$ and $\zeta'_{x_1x_2\dots x_{N-1}}$
have the same form but their coefficients, the equation 13 also
holds for the prime case, that is,
\begin{eqnarray}
& & \zeta'_{x_1x_2\dots
x_{N-1}}\otimes\Psi^-_{a_{N-1}b_{N-1}}\otimes\dots\otimes\Psi^-_{a_2b_2}\otimes
\Psi^-_{a_1b_1} = \frac{1}{2^{N-1}}\sum\limits_{i=1}^{2^{N-1}}
O_{i}\zeta'_{b_1b_2\dots b_{N-1}},
\end{eqnarray}
Then the equation 12 can be written as
\begin{eqnarray}
&& |\chi\rangle_{x_1x_2\dots x_Na_Nb_N\dots a_2b_2a_1b_1}
\nonumber \\ &=&
\frac{1}{2^{N-1}}\sum\limits_{i=1}^{2^{N-1}}(|0\rangle_{x_N}O_i\zeta_{b_1b_2\dots
b_{N-1}} + |1\rangle_{x_N}O_i\zeta'_{b_1b_2\dots b_{N-1}}) \otimes\Psi^-_{a_Nb_N} \nonumber \\
&=& \frac{1}{2^N}\sum\limits_{i=1}^{2^{N-1}}[
-|\Psi^-\rangle_{x_Na_N}(O_i\zeta_{b_1b_2\dots
b_{N-1}}|0\rangle_{b_N}+O_i\zeta'_{b_1b_2\dots
b_{N-1}}|1\rangle_{b_N}) \nonumber \\ &&
+|\Psi^+\rangle_{x_Na_N}(-O_i\zeta_{b_1b_2\dots
b_{N-1}}|0\rangle_{b_N}+O_i\zeta'_{b_1b_2\dots
b_{N-1}}|1\rangle_{b_N}) \nonumber \\ &&
+|\Phi^-\rangle_{x_Na_N}(O_i\zeta_{b_1b_2\dots
b_{N-1}}|1\rangle_{b_N}+O_i\zeta'_{b_1b_2\dots
b_{N-1}}|0\rangle_{b_N}) \nonumber \\ &&
+|\Phi^+\rangle_{x_Na_N}(O_i\zeta_{b_1b_2\dots
b_{N-1}}|1\rangle_{b_N}-O_i\zeta'_{b_1b_2\dots
b_{N-1}}|0\rangle_{b_N})]
 \nonumber \\ &=&
\frac{1}{2^N}\sum\limits_{i=1}^{2^{N-1}}(
-|\Psi^-\rangle_{x_Na_N}O_i
  + |\Psi^+\rangle_{x_Na_N}O_i\sigma_{b_N}^z +|\Phi^-\rangle_{x_Na_N}O_i\sigma_{b_N}^x
\nonumber \\ &&
+|\Phi^-\rangle_{x_Na_N}O_i\sigma_{b_N}^x\sigma_{b_N}^z)
|\xi\rangle_{b_1b_2\dots b_N}.
%\nonumber \\ &=&
%\frac{1}{2^N}\sum\limits_{i=1}^{2^N}{\cal A}_{i;x_Na_N}{\cal
%H}_{i;x_{N-1}a_{N-1}}\dots {\cal P} _{i;x_2a_2}{\cal Q}_{i;x_1a_1}
%U_{i;b_Nb_{N-1}\dots b_2b_1}\xi_{b_1b_2\dots b_{N-1}b_N}.
\end{eqnarray}
The equation 15 has shown that, if Alice performs $N$ Bell state
measurements and publishes $2N$ bits information, then conditioned
on Alice's information, Bob can recover the arbitrary state
$|\xi\rangle$ by performing at most $2N$ single-qubit operations.
As a matter of fact, in section 2 I have already shown that any
arbitrary 2-qubit state can be teleported by using and identifying
only Bell states, hence, in terms of recurrence it is easily
concluded that any arbitrary $N(N\ge3)$-qubit state can also be
successfully teleported by using and identifying only Bell states.
So far I have generalized the previous specific 2-qubit quantum
state teleportation scheme to a $N$-qubit case.

Now let us simply compare the present scheme with the counterpart
scheme in Ref.[16] as follows: (1) In the scheme in Ref.[16],
Alice needs to share a $2N$-qubit generalized Bell states with
Bob, while in this scheme Alice only needs to share $N$ same Bell
states with Bob. (2) In the scheme in Ref.[16], Alice needs to
realize a $2N$-qubit generalized Bell state measurement. In the
present scheme, only $N$ Bell states need to be identified. (3)
Alice needs to publish a messsge of $2N$ classical bits
corresponding to her $2N$-qubit generalized Bell state measurement
outcome. In the present scheme, Alice also needs to publish a
messsge of $2N$ classical bits corresponding to her $N$ Bell state
measurement outcomes. (4) In both scheme, Bob needs to perform at
most $2N$ single-qubit operations conditioned on Alice's message.
Obviously, the present scheme is preponderant.\\

{\bf 4. Summary} \\

I have explicitly shown a teleportation scheme that allows Alice
to faithfully and deterministically teleport an arbitrary 2-qubit
state to Bob. In the scheme, only two same Bell states need to be
shared by Alice and Bob. Only four Bell states must be
discriminated by Alice. Bob needs to perform at most 4
single-qubit operations conditioned on Alice's 4-bit classical
message. I have generalized this scheme to a $N(N\ge 3)$-qubit
teleportation case, where only $N$ same Bell states must be
employed and only $N$ Bell states should be identified after
quantum swapping. Bob needs to perform at most $2N$ single-qubit
operations conditioned on Alice's $2N$-bit classical message. This
leads to an important conclusion that for {\it any} qubit state
teleportation the use and identification of Bell states and
single-qubit operations are sufficient and economical. The
Comparison with the newest relevant work [16] is made, the present
schemes greatly increase the use efficiency of Bell-state resource
and decrease the experimental realization difficulty. \\

\noindent {\bf Acknowledgement} \\

This work is supported by the National Natural Science Foundation
of China under Grant No. 10304022. \\

\noindent {\bf References}\\

\noindent[1] C. H. Bennett, G. Brassard C. Crepeau,  R. Jozsa, A.
Peres and W. K. Wotters, Phys. Rev. Lett. {\bf70}, 1895 (1993).

\noindent[2] D. Bouwmeester, J. -W. Pan, K. Martle, M. Eibl, H.
Weinfurter, and A. Zeilinger, Nature (London), {\bf 390}, 575
(1997).

\noindent[3] A. Fuusawa, J. L. Sorensen, S. L. Braunstein, C. A.
Fuchs, H. J. Kimble, and E. S. Polzik, Science {\bf 282}, 706
(1998).

\noindent[4] M. A. Nilson, E. Knill, and R. Laflamme, Nature
(London), {\bf 396}, 52 (1998).

\noindent[5] M. Ikram, S. Y-. Zhu, and M. S. Zubairy, Phys. Rev. A
{\bf 62}, 022307 (2000).

\noindent[6] C. P. Yang and G. C. Guo, Chin. Phys. Lett. {\bf 16},
628 (2000).

\noindent[7] W. Son, J. Lee, M. S. Kim, and Y. -J. Park, Phys.
Rev. A {\bf 64}, 064304 (2001).

\noindent[8] J. Lee, H. Min, and S. D. Oh, Phys. Rev. A {\bf 64},
014302 (2001).

\noindent[9] J. Lee, H. Min, and S. D. Oh, Phys. Rev. A {\bf 66},
052318 (2002).

\noindent[10] T. J. Johnson, S. D. Bartlett, and B. C. Sanders,
Phys. Rev. A {\bf 66}, 042326 (2002).

\noindent[11] W. P. Bowen, N. Treps, B. C. Buchler, R. Schnabel,
T. C. Ralph, Hans-A. Bachor, T. Symul, and P. K. Lam, Phys. Rev. A
{\bf 67}, 032302 (2003).

\noindent[12] J. Fang, Y. Lin, S. Zhu, and X. Chen, Phys. Rev. A
{\bf 67}, 014305 (2003).

\noindent[13] N. Ba An, Phys. Rev. A {\bf 68}, 022321 (2003).

\noindent[14] M. Fuji, Phys. Rev. A {\bf 68}, 050302 (2003).

\noindent[15] T. Gao, Z. -x. Wang, and F. -l. Yan, Chin. Phys.
Lett. {\bf 20}, 2094 (2003).

\noindent[16] G. Rigolin, Phys. Rev. A {\bf 71}, 032303 (2005).

\noindent[17] Zhan-jun Zhang, Yong Li and Zhong-xiao Man,  Phys.
Rev. A {\bf 71}, 044301 (2005).

\noindent[18] Z. J. Zhang, J. Yang, Z. X. Man and Y. Li, Eur.
Phys. J. D. {\bf 33} 133 (2005).

\noindent[19] Z. Zhao, Y. A. Chen, A. N. Zhang, T. Yang, H. J.
Briegel and J. W. Pan, Nature (London) {\bf 430}, 54 (2004).

\end{document}